\documentclass[twocolumn,showpacs,superscriptaddress,preprintnumbers,amsmath,amssymb]{revtex4}

\usepackage{graphicx}
\usepackage{color}
\usepackage{dcolumn}
\usepackage{bm}

\begin{document}


\title{Thermodynamic scaling of dynamics in polymer melts: Predictions from the generalized entropy theory}

\author{Wen-Sheng Xu}
\affiliation{James Franck Institute, The University of Chicago, Chicago, Illinois 60637, USA}

\author{Karl F. Freed}
\email{freed@uchicago.edu}
\affiliation{James Franck Institute, The University of Chicago, Chicago, Illinois 60637, USA}
\affiliation{Department of Chemistry, The University of Chicago, Chicago, Illinois 60637, USA}



\date{\today}

\begin{abstract}
Many glass-forming fluids exhibit a remarkable thermodynamic scaling in which dynamic properties, such as the viscosity, the relaxation time, and the diffusion constant, can be described under different thermodynamic conditions in terms of a unique scaling function of the ratio $\rho^\gamma/T$, where $\rho$ is the density, $T$ is the temperature, and $\gamma$ is a material dependent constant. Interest in the scaling is also heightened because the exponent $\gamma$ enters prominently into considerations of the relative contributions to the dynamics from pressure effects (e.g., activation barriers) vs. volume effects (e.g., free volume). Although this scaling is clearly of great practical use, a molecular understanding of the scaling remains elusive. Providing this molecular understanding would greatly enhance the utility of the empirically observed scaling in assisting the rational design of materials by describing how controllable molecular factors, such as monomer structures, interactions, flexibility, etc., influence the scaling exponent $\gamma$ and, hence, the dynamics. Given the successes of the generalized entropy theory in elucidating the influence of molecular details on the universal properties of glass-forming polymers, this theory is extended here to investigate the  thermodynamic scaling in polymer melts. The predictions of theory are in accord with the appearance of thermodynamic scaling for pressures not in excess of $\sim50$ MPa. (The failure at higher pressures arises due to inherent limitations of a lattice model.) In line with arguments relating the magnitude of $\gamma$ to the steepness of the repulsive part of the intermolecular potential, the abrupt, square-well nature of the lattice model interactions lead, as expected, to much larger values of the scaling exponent. Nevertheless, the theory is employed to study how individual molecular parameters affect the scaling exponent in order to extract a molecular understanding of the information content contained in the exponent. The chain rigidity, cohesive energy, chain length, and the side group length are all found to significantly affect the magnitude of the scaling exponent, and the computed trends agree well with available experiments. The variations of $\gamma$ with these molecular parameters are explained by establishing a correlation between the computed molecular dependence of the scaling exponent and the fragility. Thus, the efficiency of packing the polymers is established as the universal physical mechanism determining both the fragility and the scaling exponent $\gamma$. 
\end{abstract}


\maketitle

\section{Introduction}

The description of the properties of glass-forming liquids poses numerous conceptual and technical problems.~\cite{JCPReview1, JCPReview2} The properties of supercooled liquids obviously vary strongly with thermodynamic state, which for a one-component system corresponds to systems with specified temperature $T$ and pressure $P$ (convenient when cooling at constant pressure) or specified temperature $T$ and density $\rho$ (convenient when cooling at constant volume). Since separate variations in density $\rho$ or temperature $T$ greatly affect the dramatic slowing down of dynamics of glassy materials, the observation that dynamical properties, such as the viscosity $\eta$, the structural relaxation time $\tau$, and the diffusion constant $D$, depend only on a single control parameter, $\rho^\gamma/T$, where $\gamma$ depends on the material, is a quite remarkable finding that is important in guiding the design of new materials. Moreover, an explanation of the origins of this thermodynamic scaling is then crucial for deepening our understanding of universal characteristics of glass formation.~\cite{Tolle0, Roland0, Floudas0, RolandMacro2010}The exponent $\gamma$ of the thermodynamic scaling also provides a measure of the relative importance of the density and temperature in controlling glassy dynamics. Hence, not surprisingly, a large body of experiments~\cite{DreyfusPRE2003, RolandPRE2004, Tarjus1, Tarjus2, RolandJCP2005, RolandPRE2005, RolandJCP2006} and simulations~\cite{CoslovichJPCB2008, CoslovichJNCS2011, CoslovichJCP2009, NgaiJPCB2010, BerthierPRL2009, BerthierJCP2011, PedersenPRL2010} have probed the nature of thermodynamic scaling of glass-forming liquids since the first observation by T\"{o}lle \textit{et al.} for \textit{ortho}-terphenyl.~\cite{Tolle1} The existence of thermodynamic scaling is well established for as diverse materials as van der Waals liquids, polymers, ionic liquids, weakly hydrogen-bonded systems, etc.~\cite{Roland0, Floudas0, RolandMacro2010}

Thermodynamic scaling in glass-forming liquids is intriguing for both fundamental and practical reasons. Probing the origin of thermodynamic scaling assists attempts to establish a universal description of the dynamics of glassy materials and to improve significantly the existing models and theories. The study is also valuable since the dynamical properties at varying thermodynamic conditions can be predicted from only a few measurements if the material conforms to the thermodynamic scaling. While some have argued that thermodynamic scaling contains little physical content,~\cite{Tarjus1, Tarjus2} the opposite wide belief is that thermodynamic scaling stems from the nature of the repulsive portion of the intermolecular potential. In fact, the thermodynamic scaling exactly holds for a system interacting with an inverse power law (IPL) potential where $\gamma$ is indeed determined by the exponent of the power law.~\cite{IPL} Strong support for this interpretation emerges from computer simulations using generalized Lennard-Jones (LJ) potentials with varying steepness~\cite{CoslovichJPCB2008, CoslovichJNCS2011} that exhibit $\gamma$ as increasing with the steepness of the potential. The study of thermodynamic scaling has  triggered the proposal of several novel concepts by Dyre and coworkers, such as that of ``strongly correlating liquids'' (i.e., liquids with strong correlations between  equilibrium fluctuations of the potential energy and the virial in a canonical ensemble) and ``isomorphs'' (i.e., curves in the phase diagram along which structure, dynamics, and some thermodynamic properties are invariant in reduced units) .~\cite{Dyre1, Dyre2, Dyre3, Dyre4, Dyre5, Dyre6} The isomorph theory indicates that the scaling exponent $\gamma$ is not constant, but depends on density,~\cite{Dyre5, Dyre6, Dyre7, Dyre8} and the power-law form $\rho^\gamma$ is only a special case. However, the power-law density scaling is a useful approximation to the isomorph scaling since a number of experiments find  a weak dependence of $\gamma$ on density under certain thermodynamic conditions~\cite{Roland0, Floudas0, RolandMacro2010} and since using a constant value of $\gamma$ leads to collapse onto a master curve of the dynamics in glass-forming liquids. Furthermore, the Avramov model, which is frequently employed in describing viscosity and relaxation data of glass-forming liquids in the $T$-$P$ plane,~\cite{Avramov} has been explored by Casalini \textit{et al.}~\cite{RolandAvramov} to rationalize the thermodynamic scaling and to show that the scaling exponent $\gamma$ is related to the Gr\"{u}neisen constant. In addition, Floudas \textit{et al.}~\cite{FloudasJCP2006} note that the monomer volume and local packing play an important role in understanding the thermodynamic scaling of polymers. Recently,  Ngai \textit{et al.}~\cite{NgaiJCP2012} present evidence that thermodynamic scaling of the $\alpha$-relaxation time stems from the properties of the Johari-Goldstein $\beta$-relaxation or the primitive relaxation of the coupling model.  

Despite intense recent efforts, understanding of the origins of thermodynamic scaling in glass-forming liquids is far from complete, and, in particular, an explanation of how molecular details of the polymers affect the thermodynamic scaling remains elusive. Although the thermodynamic scaling is a universal property that is common to both small molecules and polymers, a better understanding of its origins benefits from exploring the special features of polymer systems, such as the large changes in properties that may be produced by small alterations in molecular parameters. While dielectric spectroscopy experiments~\cite{CasaliniJCP2007} and molecular dynamics simulations~\cite{LJC} have probed the dependence of thermodynamic scaling on chain length for poly(methylmethacrylate) (PMMA) and Lennard-Jones chains, respectively, it is in general difficult experimentally to finely tune a single molecular parameter, and simulations are restricted to systems far above the glass transition temperature. Moreover, some striking empirical correlations between the scaling exponent and the fragility parameter~\cite{RolandPRE2004, RolandPRE2005, CasaliniJNCS2006, GrzybowskiPRE2006} also invite theoretical considerations. We address the above issues by employing the generalized entropy theory,~\cite{FreedACR2011, JacekACP2008, JacekJPCB2005a, JacekJPCB2005b, JacekJCP2005, JacekJCP2006, JackJCP2006, Evgeny} which has been previously developed to describe both the universal characteristics of glass formation and the specific properties of polymeric glass-forming liquids. 

One of the strengths of the generalized entropy theory is the predictive ability to treat the influence of the short-range correlations imparted by chain connectivity, semiflexibility and molecular details on the glass formation of polymers. The previously successful applications~\cite{FreedACR2011, JacekACP2008} suggest that the theory could be utilized to investigate more specific details, such as the thermodynamic scaling of polymer melts. Hence, we assess the influence of molecular details on thermodynamic scaling of polymer melts within this molecular theory of polymer glass formation. All of the following factors, including the chain rigidity, the cohesive energy, the chain length, and the length of the side groups, are demonstrated as significantly influencing the scaling exponent $\gamma$. The computed trends are found to agree well with available experiments, although the highly steep character of the square well interactions of a lattice model leads to very large values of $\gamma$ in accord with expectations and implies that only general trends should be considered. Correlations between the scaling exponent and the fragility parameter emerge from our calculations and thus suggest a universal physical mechanism for both the thermodynamic scaling of polymer dynamics and the fragility of glass-forming  polymers.

Section II begins with a brief review of the generalized entropy theory of polymer glass formation followed by a demonstration that thermodynamic scaling is well recovered within the generalized entropy theory, provided the employed pressure is not very high. Section III examines the influence of chain rigidity, cohesive energy, the lengths of the chain backbone and the side groups on thermodynamic scaling. We discuss the correlations between the scaling exponent and the fragility parameters as well as the anticipated differences from experiment caused by the very abrupt nature of the repulsive potential inherent in lattice models.

\section{Thermodynamic scaling within the generalized entropy theory}

This section begins with a brief introduction to the generalized entropy theory of polymer glass formation and a summary of the basic features of the theory that are required for the present work. We then determine whether and under what conditions the thermodynamic scaling of relaxation times emerges from the generalized entropy theory.

\subsection{Generalized entropy theory}

The generalized entropy theory has been developed with the initial aim of providing a molecular understanding of the universal properties of glass-forming polymer fluids and of explaining the strong variations in glass transition temperature $T_g$ and in fragility between different polymer materials. To this end,  the configurational entropy, a central quantity in the theoretical descriptions of the glass transition, is evaluated from the lattice cluster theory (LCT) for semiflexible polymer chains.~\cite{LCT1, LCT2} The LCT describes the influence of molecular details by employing an extended lattice model in which monomers have internal structure reflecting their size, shape and bonding patterns. By combining the LCT with the Adam-Gibbs (AG) model,~\cite{AG} the generalized entropy theory allows describing the influence of short-range correlations, induced by chain rigidity and monomer structure, on the glass formation of polymers. Note that the theory identifies the entropy density $s_c$, i.e., the configurational entropy per lattice site, as the essential quantity for use in the AG model. Thus, LCT computations of $s_c(T)$ have enabled the direct determination of three characteristic temperatures of glass formation,~\cite{JacekACP2008, JacekJPCB2005a, JacekJPCB2005b, JacekJCP2005, JacekJCP2006} namely, the ``ideal'' glass transition temperature $T_0$ where $s_c$ extrapolates to zero, the onset temperature $T_A$ which signals the onset of non-Arrhenius behavior of the relaxation time and which is evaluated from the maximum in $s_c(T)$, and the crossover temperature $T_I$ which separates two temperature regimes with qualitatively different dependences of the relaxation time on temperature and which is evaluated from the inflection point in $Ts_c(T)$. The relaxation time $\tau$ is then obtained from $s_c$ and the AG relation,
\begin{equation}
\tau=\tau_\infty\exp[\beta\Delta\mu s_c^\ast/s_c(T)],
\end{equation}
where $\beta=1/(k_BT)$ with $k_B$ Boltzmann's constant, $\tau_\infty$ is the high temperature limit of the relaxation time, $\Delta\mu$ is the limiting temperature independent activation energy at high temperatures, and $s_c^\ast$ is the high temperature limit of $s_c(T)$. The high temperature activation energy is estimated  from the empirical relation $\Delta\mu=6k_BT_I$ and $ \tau_\infty$ is set to be $10^{-13}$ s. Thus, the structural relaxation time is computed within the generalized entropy theory without adjustable parameters beyond those used in the LCT for the thermodynamics of polymers. The temperature dependence of the relaxation time enables determination of the glass transition temperature $T_g$, the fourth characteristic temperature, from the common empirical definition $\tau(T_g)=100$ s. (The use of a Lindermann criterion yields essentially equivalent results.~\cite{JacekJPCB2005a}) Therefore, glass formation is described by the generalized entropy theory as a broad thermodynamic transition with four characteristic temperatures ($T_A$, $T_I$, $T_g$, $T_0$). 

The generalized entropy theory has previously been applied to investigate essential molecular and physical features affecting glass-formation, ~\cite{JacekACP2008, JacekJPCB2005a, JacekJPCB2005b, JacekJCP2005, JacekJCP2006, JackJCP2006} and recently to model the glass formation of poly($\alpha$-olefins).~\cite{Evgeny} These studies explain some experimentally observed trends, while many theoretical predictions are confirmed by experiments~\cite{FreedACR2011, JacekACP2008}, suggesting that the theory correctly captures general trends of glass formation in polymers.  We further extend the theory in this paper to investigate thermodynamic scaling of the dynamics of polymer melts, whether the LCT also predicts thermodynamic scaling, how the exponent $\gamma$ varies with molecular parameters, and the relation of $\gamma$ to other measurable quantities. The theory describes the polymers as involving monomers that extend over several lattice sites with connectivities corresponding to the molecular structures. The key parameters of the theory and their significance are as follows: (1) The bending energy $E_b$ describes the chain rigidity which may differ between the backbone and the side groups. The bonds are fully flexible when $E_b=0$, whereas they are completely rigid in the $E_b\rightarrow\infty$ limit. (2) The cohesive energy parameter $\epsilon$ describes the net attractive van der Waals interactions between the nearest neighbor united atom groups. (3) The monomer structure and molecular weight enter into the theory through a set of geometrical indices, providing a substantial improvement over the traditional lattice theories. Further information on the generalized entropy theory can be found in recent reviews~\cite{FreedACR2011, JacekACP2008}.

\subsection{Testing thermodynamic scaling}

\begin{figure}[tb]
 \centering
 \includegraphics[angle=0,width=0.48\textwidth]{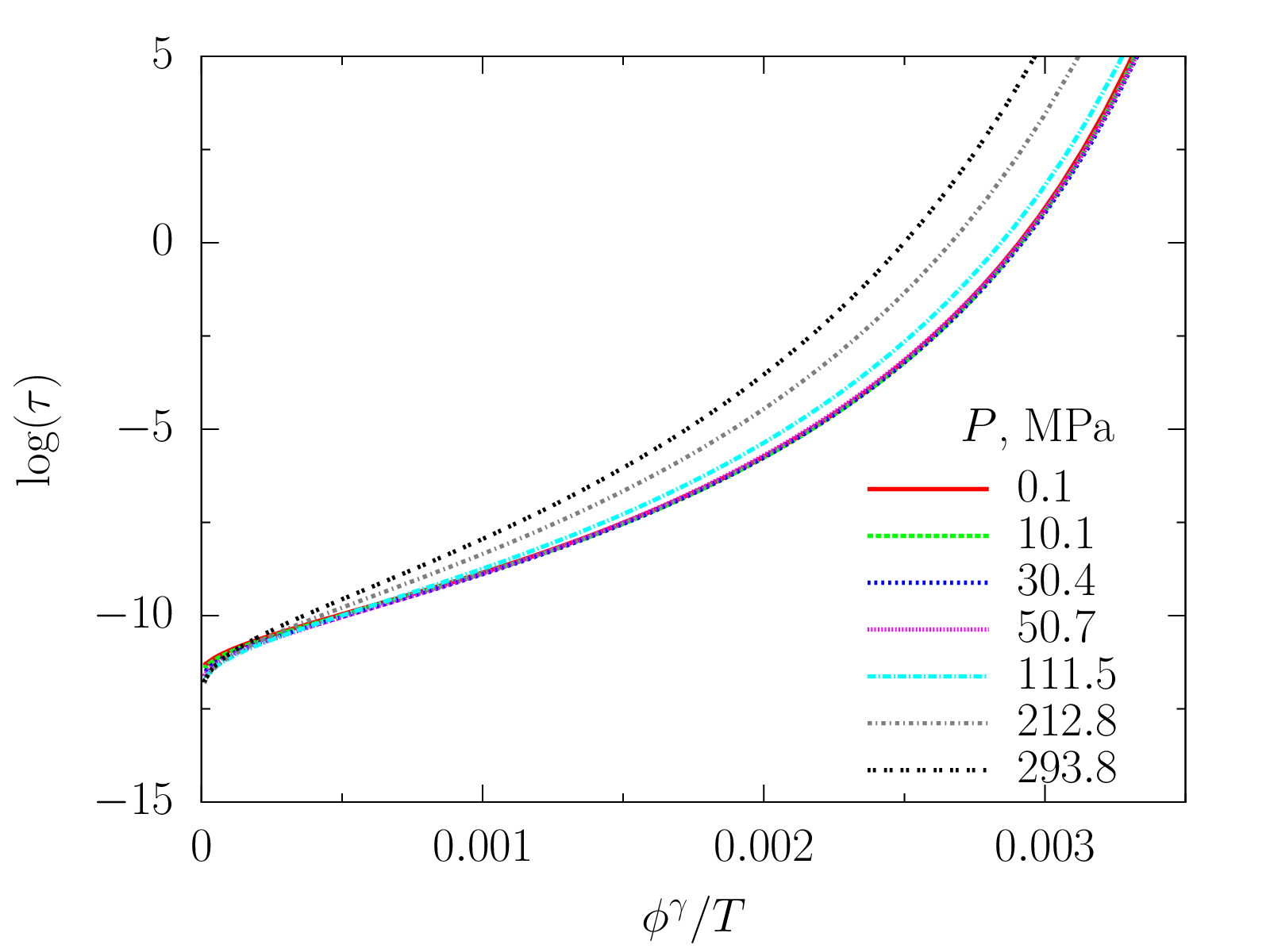}
 \caption{The logarithm of relaxation time $\log(\tau)$ calculated from the generalized entropy theory as a function of the ratio $\phi^\gamma/T$ with $\gamma=14.0$ for various pressures. The computations are performed for a melt of chains with the structure of poly(propylene) (PP) with $z=6$, $a_{cell}=2.7$ \AA{},  $\epsilon/k_B=200$ K, $E_b/k_B=400$ K and $N=8000$. The same values of $z$, $a_{cell}$, $\epsilon$, $E_b$ and $N$ are used in the computations presented in Fig. 2. $\tau$ is given in units of seconds, which is also used in Fig. 4. Thermodynamic scaling holds well for low pressures, but relaxation times for $P\gtrsim 50$ MPa deviate from the master curve.}
\end{figure}

\begin{figure}[t]
 \centering
 \includegraphics[angle=0,width=0.48\textwidth]{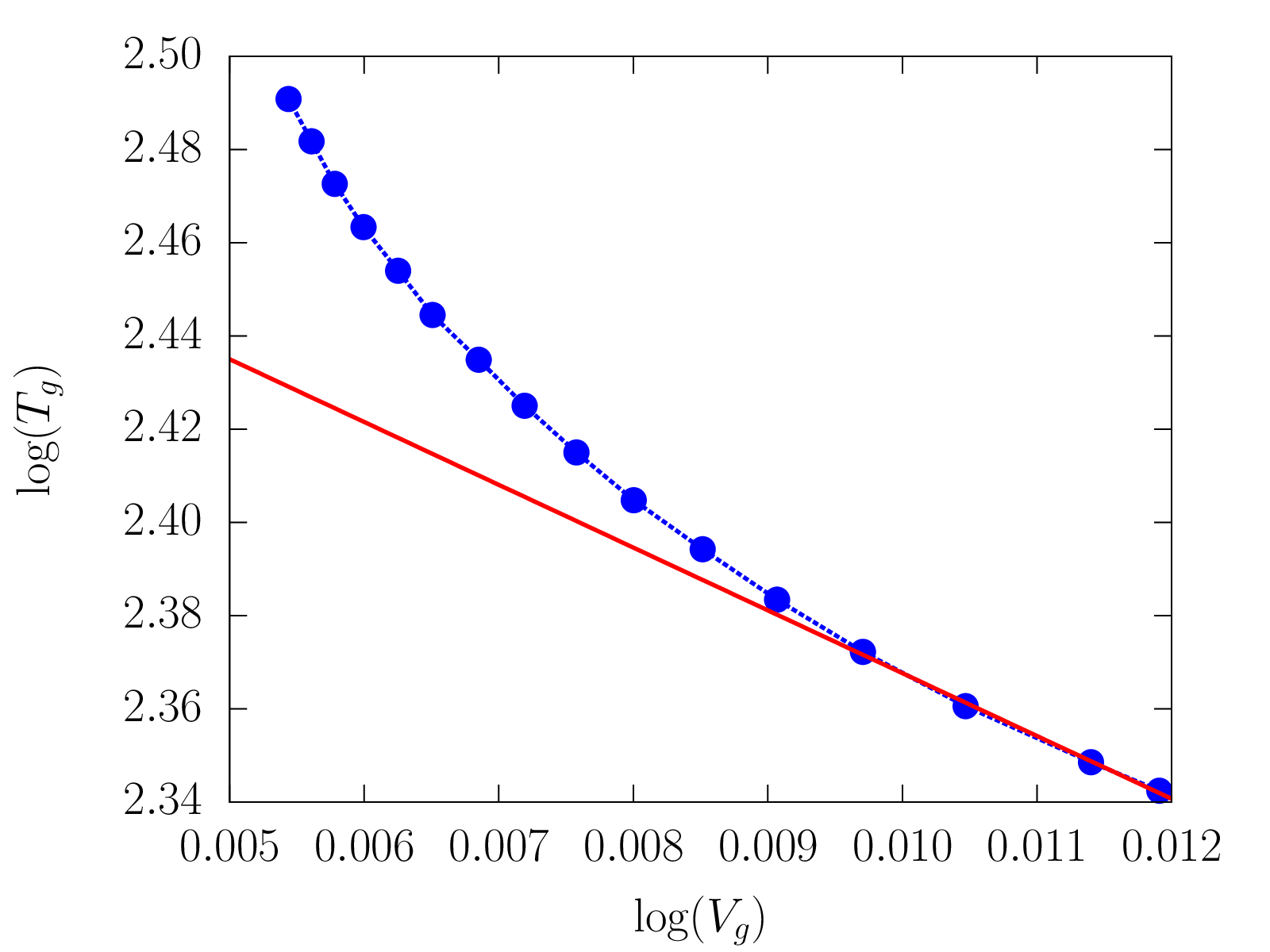}
 \caption{The logarithm of glass transition temperature $\log(T_g)$ as a function of the logarithm of glass transition specific volume $\log(V_g)$. The red solid line is a linear fit to the data of slope $13.47$ for $P \leq 50.7$ MPa. A linear relationship between $\log(T_g)$ and $\log(V_g)$ indicates that the relaxation times obey thermodynamic scaling.}
\end{figure}

As a first test of whether the generalized entropy theory describes thermodynamic scaling and as a prelude to more extensive calculations below that probe the  variation of the scaling exponent with each of the individual parameters of the model, we choose a polymer system with the structure of poly(propylene) (PP) because such a structure requires the minimal number of parameters in the LCT. The following parameters are used in the calculations: the lattice coordination number ($z=6$),  the cell volume ($v_{cell}=a_{cell}^3=2.7^3$ \AA{}$^3$), the cohesive energy parameter ($\epsilon/k_B=200$ K), the bending energy ($E_b/k_B=400$ K) and the chain length ($N=8000$). Following  previous work, the computations are performed at constant pressure. Then, the relaxation time $\tau$ and volume fraction $\phi$ are obtained as a function of temperature for various pressures up to $P=293.8$ MPa and are used to analyze the nature of  thermodynamic scaling. Although recent work suggests the use of the reduced-unit relaxation time for analyzing the thermodynamic scaling,~\cite{RolandJCP2011} we employ the absolute values which also work well in most cases and which yield almost identical values of the scaling exponent. The scaling exponent $14.0$ collapses the data for $P\leq50.7$ MPa onto a master curve, but  Fig. 1 implies that deviations from a master curve appear when the pressure exceeds $\sim50$ MPa.

As noted in Ref.~\cite{PaluchJPCM2007}, another convenient method to estimate the scaling exponent is from the linear relationship between $\log(T_g)$ and $\log(V_g)$,
\begin{equation}
\log (T_g)=A-\gamma\log(V_g),
\end{equation}
that follows as a consequence of thermodynamic scaling, where $V_g$ is the specific volume (defined as $V=1/\phi$) at the glass transition temperature. Figure 2 displays the correlation between $\log (T_g)$ and $\log(V_g)$ that varies linearly only in the low-pressure regime. 

According to isomorph theory~\cite{Dyre5, Dyre6, Dyre7, Dyre8}, the scaling exponent is not constant, but generally is a function of density. Thus, a natural question  is whether the departures at higher pressures in our computations arise from using a constant value of $\gamma$. We have tried to employ the density-dependent scaling exponent $\gamma(\phi)$ to collapse the data presented in Fig. 1 using $\gamma(\phi)$ obtained from the slopes in Fig. 2, but  improvements are minimal at high pressures. Additionally, our computations display a similar breakdown at higher pressures for R\"{o}ssler scaling,~\cite{Rossler, JacekACP2008} in which data for the normalized relaxation times $\tau/\tau(T_I)$ collapse to a single function of $(T_g/T)(T_I-T)/(T_I-T_g)$ (data not shown). We thus believe that the departures at higher pressures are probably due to inherent limitations of the employed lattice model. Therefore, subsequent computations are performed in the low-pressure regime and probe the calculated variation of the scaling exponent with the molecular parameters of the LCT using data for four pressures ($P=0.1$, $10.1$, $30.4$, $50.7$ MPa).

\subsection{Comparison with experiment}

\begin{figure}[tb]
 \centering
 \includegraphics[angle=0,width=0.48\textwidth]{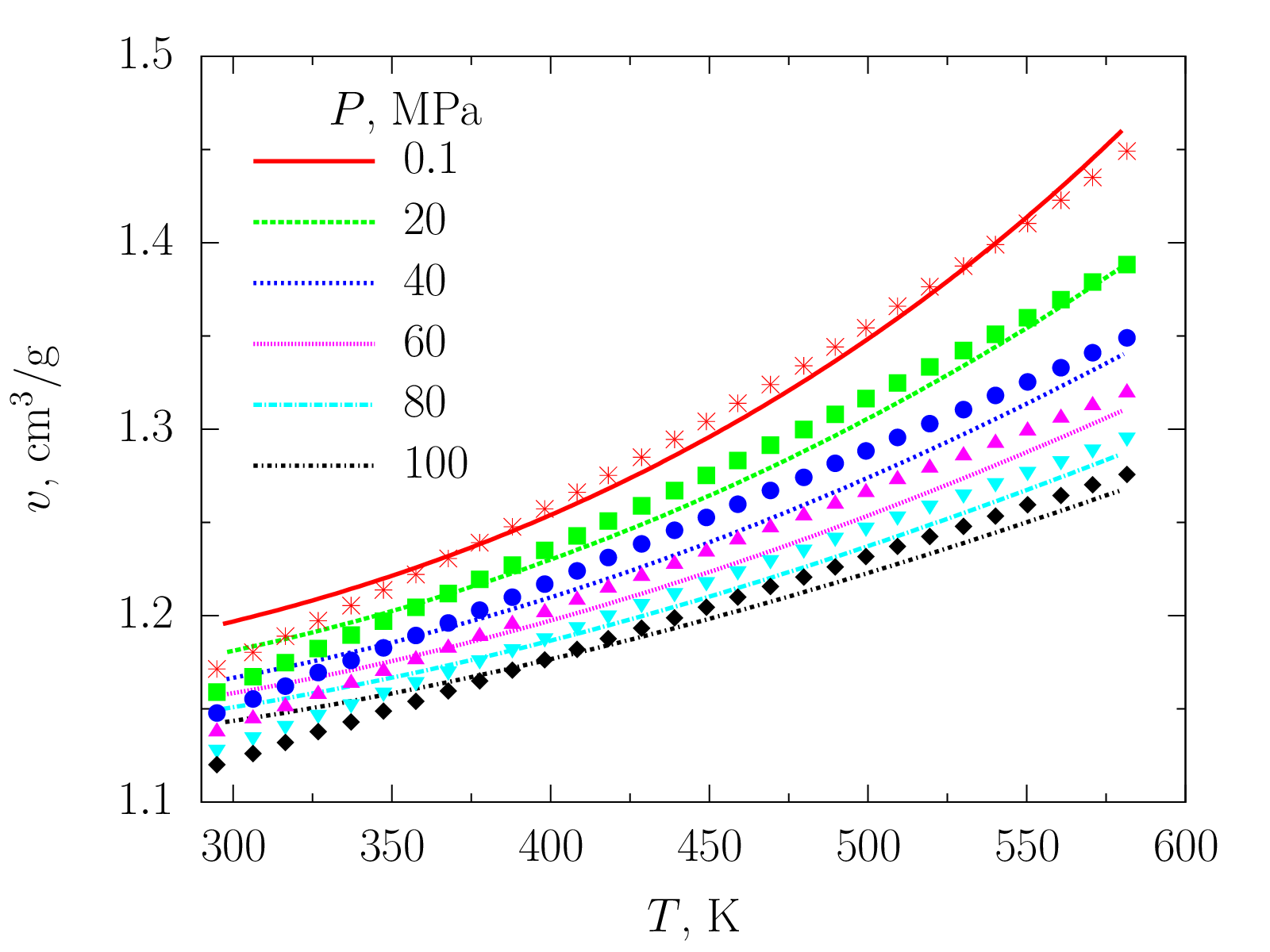}
 \caption{Specific volumes $v$ as a function of temperature $T$ for various pressures~\cite{footnote1}. The symbols are experimental data taken from Ref.~\cite{PVT} for atactic PP with high molecular weight, and the lines are results calculated from the generalized entropy theory for a melt of chains with the PP structure with $z=6$, $E_b/k_B=409$ K and $N=8000$. The cell volume parameter $a_{cell}$ and cohesive energy $\epsilon$ are adjusted to decrease with pressure in order to better describe the experimental data. The parameters are summarized in Table I.}
\end{figure}

\begin{table}[t] 
\centering
\small\addtolength{\tabcolsep}{9pt}
\caption{Cell volume parameters $a_{cell}$, cohesive energies $\epsilon$, calculated glass transition temperatures $T_g^{cal}$, and experimental glass transition temperatures $T_g^{exp}$~\cite{Pressure} for various pressures. Using the pressure dependent $a_{cell}$ and $\epsilon$ along with the bending energy $E_b/k_B=409$ K produces the calculated pressure dependence of the glass transition temperature in good agreement with experimental one.}
\begin{tabular}{@{}rrrrr@{}}
\hline\hline
$P$, MPa&$a_{cell}$, \AA{}&$\epsilon/k_B$, K&$T_g^{cal}$, K&$T_g^{exp}$, K\\
\hline
$0.1$& $3.000$ & $266$ & $258.2$ & $258.0$\\
$20$ & $2.990$ & $260$ & $262.6$ & $262.4$\\
$40$ & $2.980$ & $253$ & $266.6$ & $266.5$\\
$60$ & $2.975$ & $246$ & $270.8$ & $270.6$\\
$80$ & $2.970$ & $238$ & $274.5$ & $274.4$\\
$100$ & $2.965$ & $230$ & $278.2$ & $278.2$\\
\hline\hline
\end{tabular}
\end{table}

The above illustrative computations exhibit a much larger scaling exponent than experimentally observed for PP, where $\gamma$ is found to be around $2.0$ for atactic PP.~\cite{HollanderJNCS2001} The model parameters, $a_{cell}$ and $\epsilon$, used in these calculations are taken as independent of pressure to enable subsequent studies of how $\gamma$ varies as a single parameter is changed. However, because the experimental equation of state for PP displays the coefficient of thermal expansion and the isothermal compressibility as roughly linearly dependent on pressure, the above illustrative calculations poorly describe the equation of state of PP. The equation of state, however, may be described reasonably well when recognizing that as the pressure is elevated, the average interatomic spacings decrease, so the cell volume parameter $a_{cell}$ should diminish. Similarly, interatomic forces at elevated pressures are more repulsive on average, so the effective attractive interaction energy $\epsilon$ should decrease (be less attractive) as pressure increases. Thus, the parameters $\epsilon$ and $a_{cell}$ are chosen to obtain a reasonable fit to the equation of state data for atactic PP~\cite{PVT} and thus decrease with pressure (see Table I).  As shown in Fig. 3, the computations provide a satisfactory description of the experimental equation of state data over a reasonable range of temperatures and pressures. Deviations become evident as temperature decreases, but the overall agreement enormously improves over the treatment with pressure independent parameters. Using these parameters and a pressure independent bending energy ($E_b/k_B=409$ K), the computations perfectly reproduce the experimental data for pressure dependence of the glass transition temperature, as displayed in Table I, where the calculated glass transition temperatures $T_g^{cal}$ are presented together with the experimental glass transition temperatures $T_g^{exp}$ determined from dielectric spectroscopy experiments of Gitsas and Floudas.~\cite{Pressure} We further find that thermodynamic scaling persists for low pressures, but with the larger scaling exponent ($\gamma\approx23$ ). 

As discussed below, the computed scaling exponent generally significantly exceeds all available experimental data where $\gamma$ lies in the range from $0.13$ to $8.5$ for most materials.~\cite{Roland0, Floudas0} The qualitative argument that the magnitude of scaling exponent $\gamma$ reflects the steepness of the repulsive part of the potential suggests that our lattice model for polymers correspond to a system with a very steep potential and, therefore, that the generalized entropy theory can only capture qualitative trends for $\gamma$ of polymer melts. In fact, the lattice model is widely believed as implicitly assuming a square well potential and thus corresponds to a system described by a power law (i.e., $r^{-n}$, where $r$ is the separation between two particles) fluid with an infinite exponent $n$. The treatment of correlations (such as chain connectivity, semiflexibility and monomer molecular structure) by the lattice cluster theory presumably softens the repulsions, but the scaling exponent $n$ tends towards an infinity for a monomeric system where the effective repulsion is steepest.

\section{Results and discussion}

This section provides a systematic investigation of the influence of the bending energy, cohesive energy, backbone and side chain lengths on the thermodynamic scaling of polymer melts, with particular focus on how these molecular factors influence the scaling exponent $\gamma$ which provides an important measure reflecting the degree to which volume effects govern the temperature and pressure dependence of the relaxation times. Whereas correlations determined experimentally suffer from the fact that several molecular parameters change simultaneously as the chemical species varies, calculations may be performed varying one parameter at a time. Thus, we examine the presence of a correlation between the fragility and the scaling exponent when only a single parameter is varied. For this reason, the calculations consider the pressure independent $a_{cell}$ and $\epsilon$ to enable varying one of them independently, with the recognition that the general trends would be the same. 

\subsection{Influence of the bending energy}

\begin{figure}[tb]
 \centering
 \includegraphics[angle=0,width=0.48\textwidth]{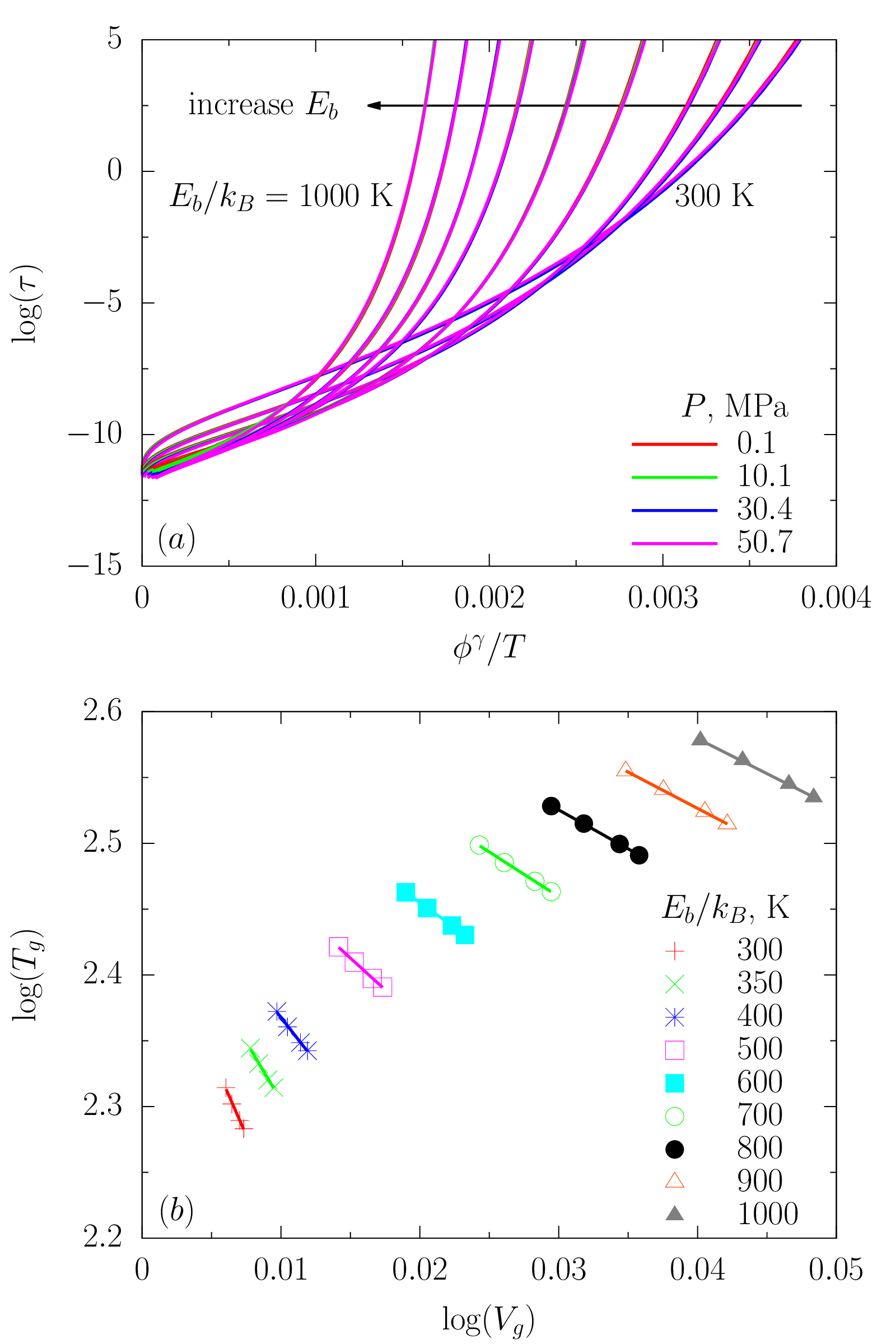}
 \caption{(a) The logarithm of relaxation time $\log(\tau)$ as a function of the ratio $\phi^\gamma/T$ for various $E_b$. (b) The logarithm of glass transition temperature $\log(T_g)$ as a function of the logarithm of glass transition specific volume $\log(V_g)$ for various $E_b$. The solid lines in (b) are linear fits. The computations are performed for the PP structure with $z=6$, $a_{cell}=2.7$ \AA{}, $\epsilon/k_B=200$ K and $N=8000$. The same values of $z$, $a_{cell}$, $\epsilon$ and $N$ are used in the computations presented in Figs. 5 and 7.}
\end{figure}

\begin{figure}[tb]
 \centering
 \includegraphics[angle=0,width=0.48\textwidth]{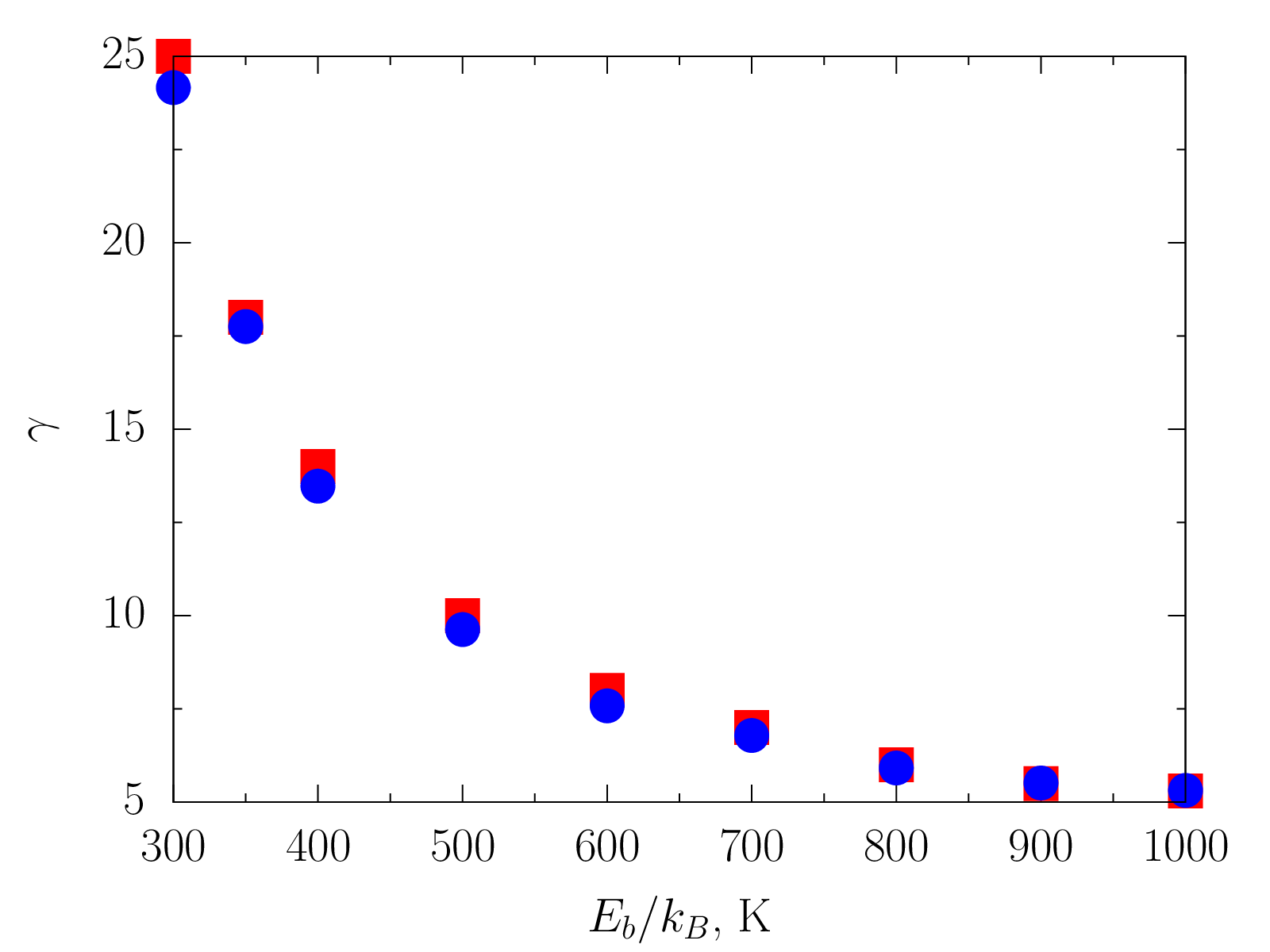}
 \caption{Scaling exponents $\gamma$, estimated from two independent methods, as a function of bending energy $E_b$. The red squares and blue circles are obtained from relaxation times and Eq. (2), respectively.}
\end{figure}

\begin{figure}[b]
 \centering
 \includegraphics[angle=0,width=0.48\textwidth]{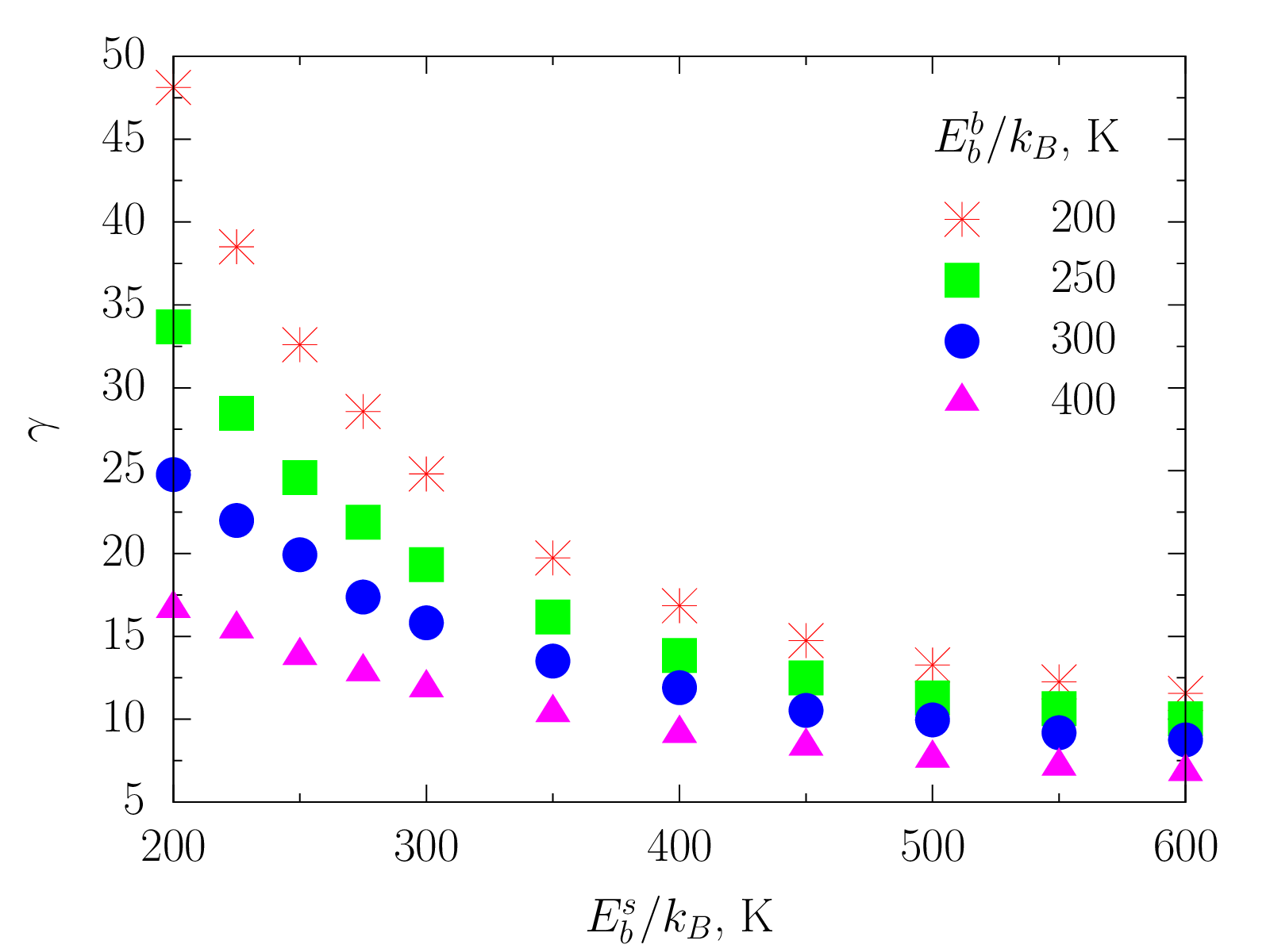}
 \caption{Scaling exponents $\gamma$ as a function of side group bending energy $E_b^s$ for various fixed backbone bending energies $E_b^b$. The calculations consider a poly($1$-pentene) (PPe) structure with $z=6$, $a_{cell}=2.7$ \AA{}, $\epsilon/k_B=200$ K and $N=8000$.}
\end{figure}

The degree of chain rigidity is modeled in terms of the bending energy $E_b$ in the generalized entropy theory and has been shown to strongly affect the fragility and the glass transition temperature.~\cite{FreedACR2011, JacekACP2008} We now consider how the bending energy influences the thermodynamic scaling for a polymer system with the PP structure because its treatment requires only a single backbone bending energy. The cell volume parameter, cohesive energy and chain length are fixed. $\tau$, $T_g$ and $V_g$ are then computed as a function of $T$ along four isobars ($P=0.1$, $10.1$, $30.4$, $50.7$ MPa). The results displayed in Fig. 4 use the two independent approaches introduced in Subsec. II (B) to estimate the scaling exponent. Both methods find scaling for pressures up to $50.7$ MPa. Figure 5 presents the computed scaling exponent as a function of bending energy from the two methods which again generate almost identical results. Thus, the calculated scaling exponent is presented using the more convenient linear relation between $\log (T_g)$ and $\log(V_g)$. 

Figure 5 clearly indicates that the scaling exponent decreases with the backbone stiffness and tends to saturate at large bending energies. While the PP structure is chosen above for illustrating the effect of the backbone bending energy, the side groups can have a separate bending energy when the length of the side groups $n$ is sufficiently large ($n\geq2$). Since the relative flexibility of the backbone and side groups has been shown to strongly correlate with the fragility of polymer glasses,~\cite{FreedACR2011, Evgeny} it is instructive to examine the influence of the side group rigidities. We thus consider a melt of polymers with the structure of poly($1$-pentene) (PPe) as an illustration since this structure has been employed by Dudowicz \textit{et al.}~\cite{JacekACP2008, JacekJPCB2005a, JacekJPCB2005b, JacekJCP2005, JacekJCP2006} to investigate the universal properties of glass formation for three different classes of polymers. We first find that the scaling exponent still decreases as the backbone bending energy grows for fixed side group bending energies (data not shown). The results displayed in Fig. 6 show that the side group bending energy has a similar effect on the scaling exponent for fixed backbone bending energies. Hence, the scaling exponent diminishes as either the backbone or side chain stiffen. Our computations thus indicate that the density effect on the dynamics of polymers is enhanced as the chain becomes flexible, a trend in good agreement with experiments; e.g., the scaling exponent is $1.25$ for PMMA,~\cite{Tarjus1} while it significantly increases for more flexible polymers such as poly(methylphenylsiloxane), where $\gamma$ is equal to $5.6$.~\cite{PaluchJCP2002}

\begin{figure}[tb]
 \centering
 \includegraphics[angle=0,width=0.48\textwidth]{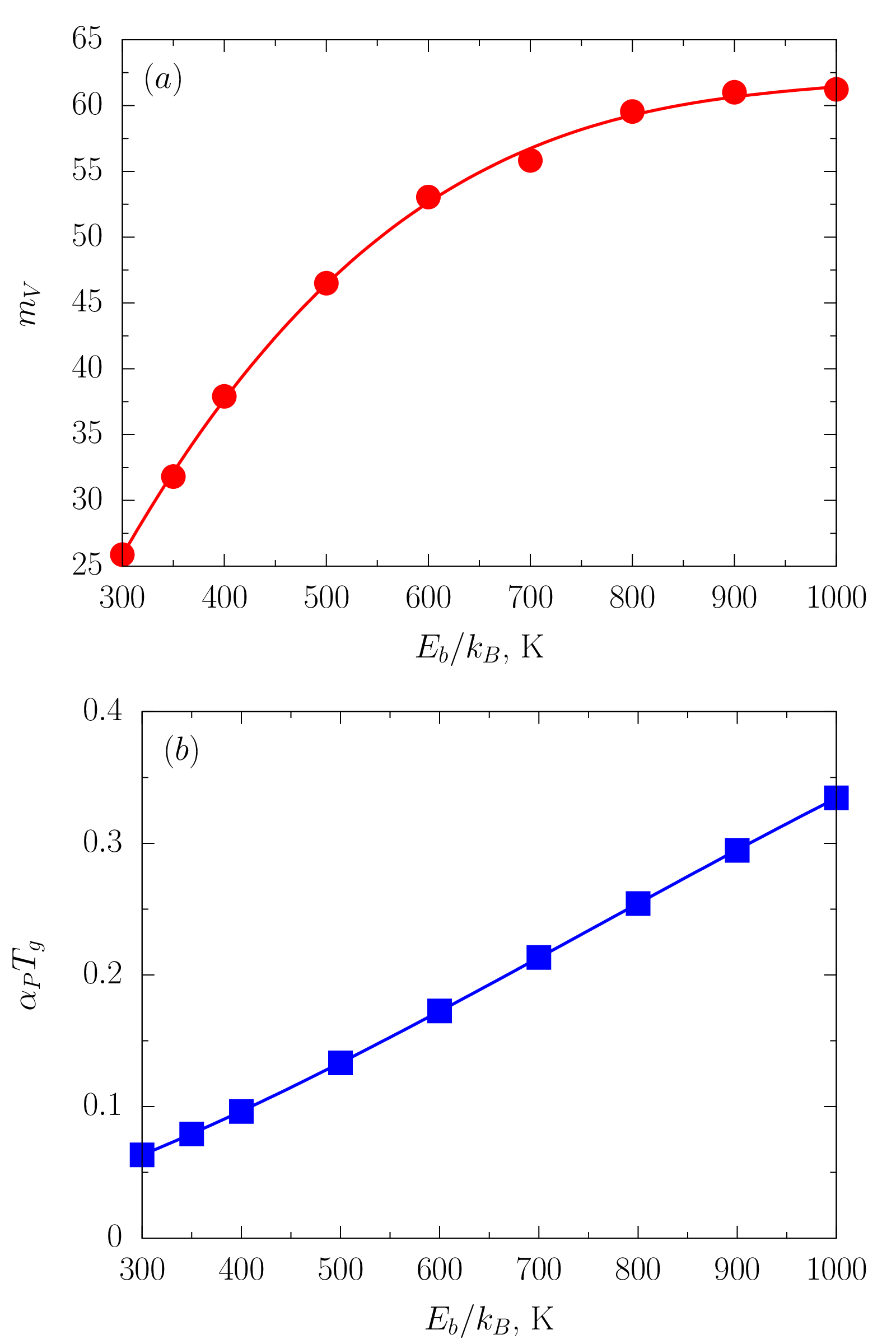}
 \caption{(a) Isochoric fragility parameter $m_V$ as a function of bending energy $E_b$. (b) The product of isobaric expansion coefficient $\alpha_P$ at the glass transition point and glass transition temperature $T_g$ as a function of bending energy $E_b$ at constant pressure ($P=0.1$ MPa). The lines are a guide to the eye.}
\end{figure}

Fragility is a basic property of glass-forming liquids, and its relation to experimental data for thermodynamic scaling has been discussed.~\cite{RolandPRE2004, RolandPRE2005, CasaliniJNCS2006, GrzybowskiPRE2006} It is straightforward to show that thermodynamic scaling implies,
\begin{equation}
\frac{m_V}{m_P}=\frac{1}{1+\gamma \alpha_P T_g},
\end{equation}
where $m_V=\partial \log (\tau)/\partial (T_g/T)|_{V, T=T_g}$ is the isochoric fragility parameter, $m_P=\partial \log (\tau)/\partial (T_g/T)|_{P, T=T_g}$ the isobaric fragility parameter and $\alpha_P=(1/V)(\partial V/\partial T)|_{P, T=T_g}$ the isobaric expansion coefficient at the glass transition temperature. Moreover, $m_V/m_P$ can be directly related to the ratio of the isochoric activation energy $E_V=R \partial \log (\tau)/\partial (T^{-1})|_V$ to the isobaric activation enthalpy $H_P=R \partial \log (\tau)/\partial (T^{-1})|_P$ with $R$ the gas constant, i.e.,
\begin{equation}
\frac{m_V}{m_P}=\left . \frac{E_V}{H_P} \right |_{T=T_g}.
\end{equation}

Since the computations are performed at constant pressure, $m_P$ and $\alpha_P$ can be directly determined according to their original definition. Then, the isochoric fragility parameter $m_V$ is calculated from Eq. (3) using the values of $m_P$ and $\alpha_P$ at $P=0.1$ MPa, or equivalently from $m_V=\partial \log (\tau)/\partial (\Gamma/\Gamma_g)|_{\Gamma=\Gamma_g}$, where $\Gamma=\phi^\gamma/T$ and the glass transition point $\Gamma_g$ is similarly defined by $\tau(\Gamma_g)=100$ s. Both methods produce identical results. As illustrated in Fig. 7 (a), larger bending energy yields increased $m_V$, as is indeed already evident from the curves in Fig. 4(a). Thus, the variation of the bending energy affects both $m_V$ and $m_P$ similarly. This result is not surprising in view of the positive correlation between $m_P$ and $m_V$ found empirically~\cite{RolandPRE2005, GrzybowskiPRE2006} and from our computations described in Subsec. III (E). The generalized entropy theory also reveals that the packing efficiency provides a general determinant of the fragility of polymers.~\cite{FreedACR2011, JacekACP2008} Thus, the density scaling for the dynamics of glassy polymers and the magnitude of the scaling exponent may also be a consequence of the packing efficiency since the strong correlation between fragility and the scaling exponent is established in Subsec. III (E).

Experiments find that $\alpha_P T_g$ is approximately constant for most glassy materials~\cite{RolandPRE2004, RolandPRE2005} so that a master curve can be constructed between $E_V/H_P$ and $\gamma$ based on Eqs. (3) and (4) for a variety of glassy materials. While such an empirical finding is very appealing, our computations indicate that it breaks down when a single molecular parameter is varied. As an illustration,  Figure 7(b) exhibits $\alpha_P T_g$ as monotonically increasing with $E_b$.  Similarly, $\alpha_P T_g$ systematically changes with other molecular parameters, such as cohesive energy, chain length and side group length, when the other parameters are held constant. Therefore, a master curve between $E_V/H_P$ and $\gamma$ cannot be established within the generalized entropy theory. It should be emphasized that empirically observed correlations between $E_V/H_P$ and $\gamma$ reflect its approximate nature.

\subsection{Influence of the cohesive energy}

\begin{figure}[tb]
 \centering
 \includegraphics[angle=0,width=0.48\textwidth]{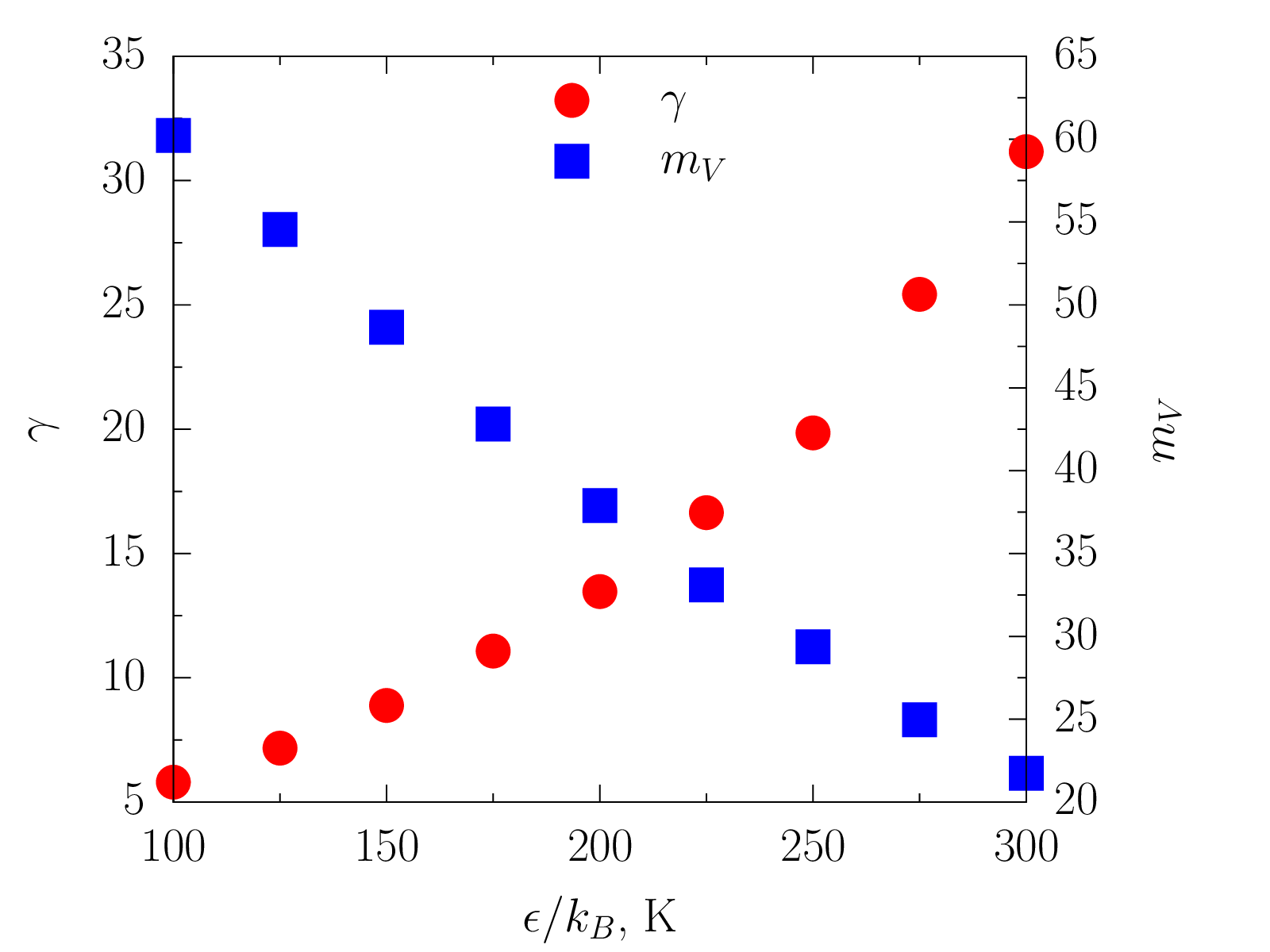}
 \caption{Scaling exponent $\gamma$ and isochoric fragility parameter $m_V$ as a function of cohesive energy $\epsilon$. The computations are performed for a polymer melt with the PP structure with $z=6$, $a_{cell}=2.7$ \AA, $E_b/k_B=400$ K and $N=8000$.}
\end{figure}

The cohesive energy parameter $\epsilon$ describes the net attractive interactions between nearest neighbor united atom groups and strongly affects the equation of state. The previous analyses of the generalized entropy theory~\cite{Evgeny} show that the glass transition temperature increases, while the fragility decreases, as the cohesive energy grows. Again, this is rationalized in the generalized entropy theory in terms of the packing efficiency of the polymers since larger attractive interactions are expected to induce the polymers to pack more efficiently. The influence of the cohesive energy parameter on the thermodynamic scaling of polymers is examined for a high molar mass polymer with the PP structure for varying cohesive energies while keeping other parameters fixed. The resulting scaling exponent $\gamma$ and isochoric fragility parameter $m_V$ are presented as a function of $\epsilon$ in Fig. 8. Not surprisingly, $\gamma$ significantly increases when $\epsilon/k_B$ grows from $100$ K to $300$ K. Thus, our results confirm the importance of the packing efficiency in determining the density-temperature scaling on the dynamics of supercooled polymer melts. Moreover, $m_V$ changes in the opposite direction, indicating again a negative correlation between the scaling exponent and the fragility.

\subsection{Influence of the chain length}

\begin{figure}[b]
 \centering
 \includegraphics[angle=0,width=0.48\textwidth]{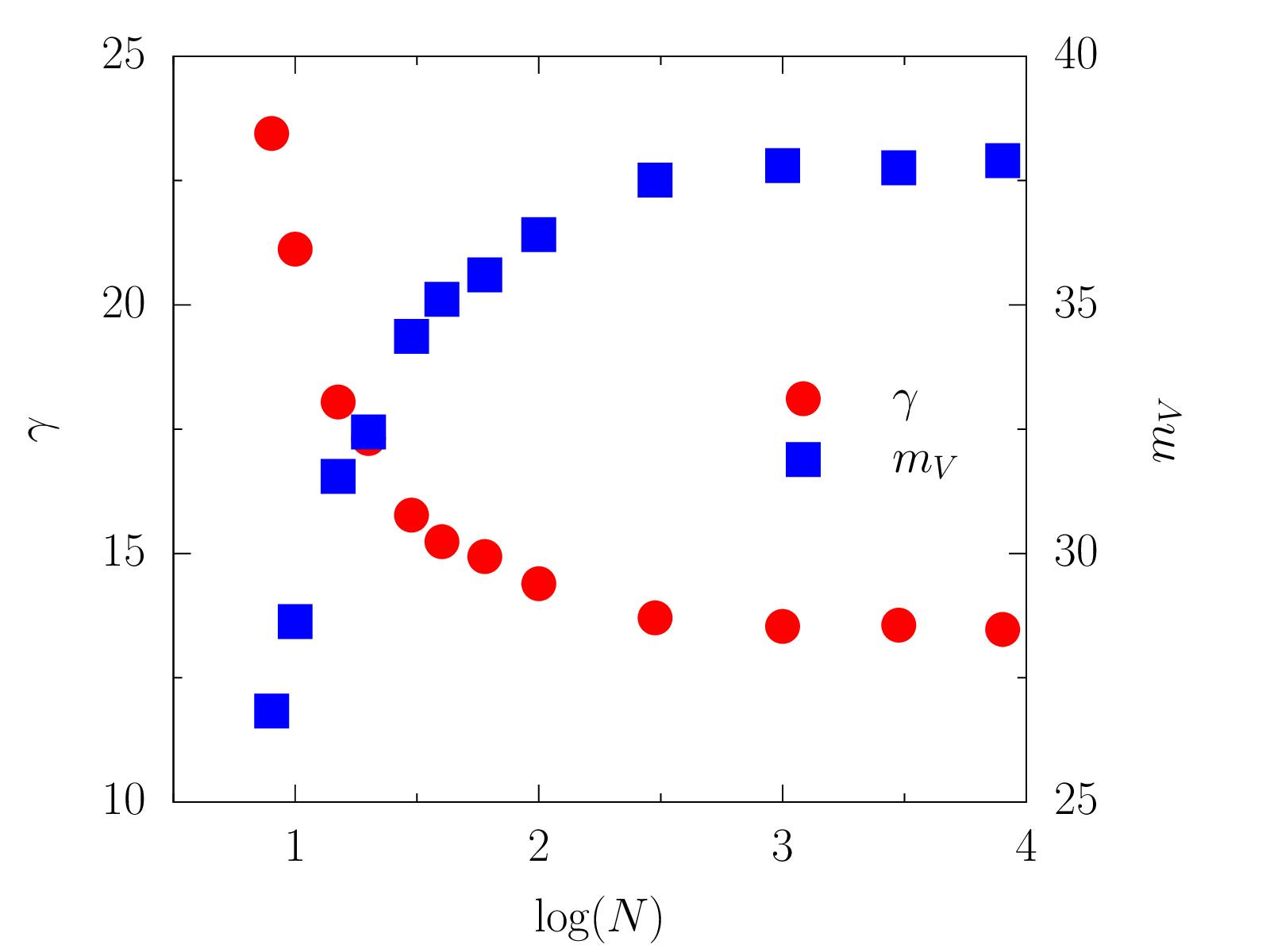}
 \caption{Scaling exponent $\gamma$ and isochoric fragility parameter $m_V$ as a function of chain length $N$. The computations are performed for a polymer melt with the PP structure with $z=6$, $a_{cell}=2.7$ \AA, $\epsilon/k_B=200$ K, $E_b/k_B=400$ K and $N=8000$. The smallest value for $N$ is $8$.}
\end{figure}

Experiments in general reveal that longer chain polymers exhibit higher glass transition temperatures and higher fragilities. Such a trend has been successfully reproduced by the generalized entropy theory.~\cite{JacekACP2008, Evgeny} These results indicate that the chain length also serves as an important molecular parameter to control the properties of glassy polymers. Recently, Casalini \textit{et al.}~\cite{CasaliniJCP2007} have experimentally detected the impact of varying chain length on thermodynamic scaling by analyzing data for the relaxation times of PMMA samples with four molecular weights. The scaling exponent decreases from $3.7$  to $1.8$ when the chain length grows from $N=3$ to $N=1500$,  while the isochoric fragility varies in the opposite direction. Thus, increasing chain length reduces the importance of the density (or volume) in determining the dynamics of polymers. Figure 9 presents the variation of the scaling exponent and the isochoric fragility parameter with chain length $N$ using calculations for a polymer melt with the PP structure for different chain lengths.  Although the generalized entropy theory can, in principle, be applied to treat monomeric systems with $N=1$,  the mean-field approximations embodied in the LCT are less faithful for such small $N$. Thus,  the computations consider $N \geq 8$. Figure 9 exhibits the scaling exponent as first quickly decreasing with increasing $N$ and then remaining nearly constant at sufficiently large $N$, while the isochoric fragility parameter again proceeds in the opposite direction. This computed trend is in accord with experimental observations described above. The predicted saturation of $\gamma$ and $m_V$ at large $N$ is physically reasonable since there must be a lower bound on $N$ above which most of the properties of polymer melts become insensitive to the change of chain length. Interestingly, recent computer simulations~\cite{LJC} indicate that the scaling exponent increases with chain length for a flexible LJ chain model. However, such a simple model probably does not account for the complicated effects in polymers and may not capture the correct trends for a real polymer material. Instead, our computations provide theoretical agreement for the experimental findings.

\subsection{Influence of the side group length}

\begin{figure}[tb]
 \centering
 \includegraphics[angle=0,width=0.48\textwidth]{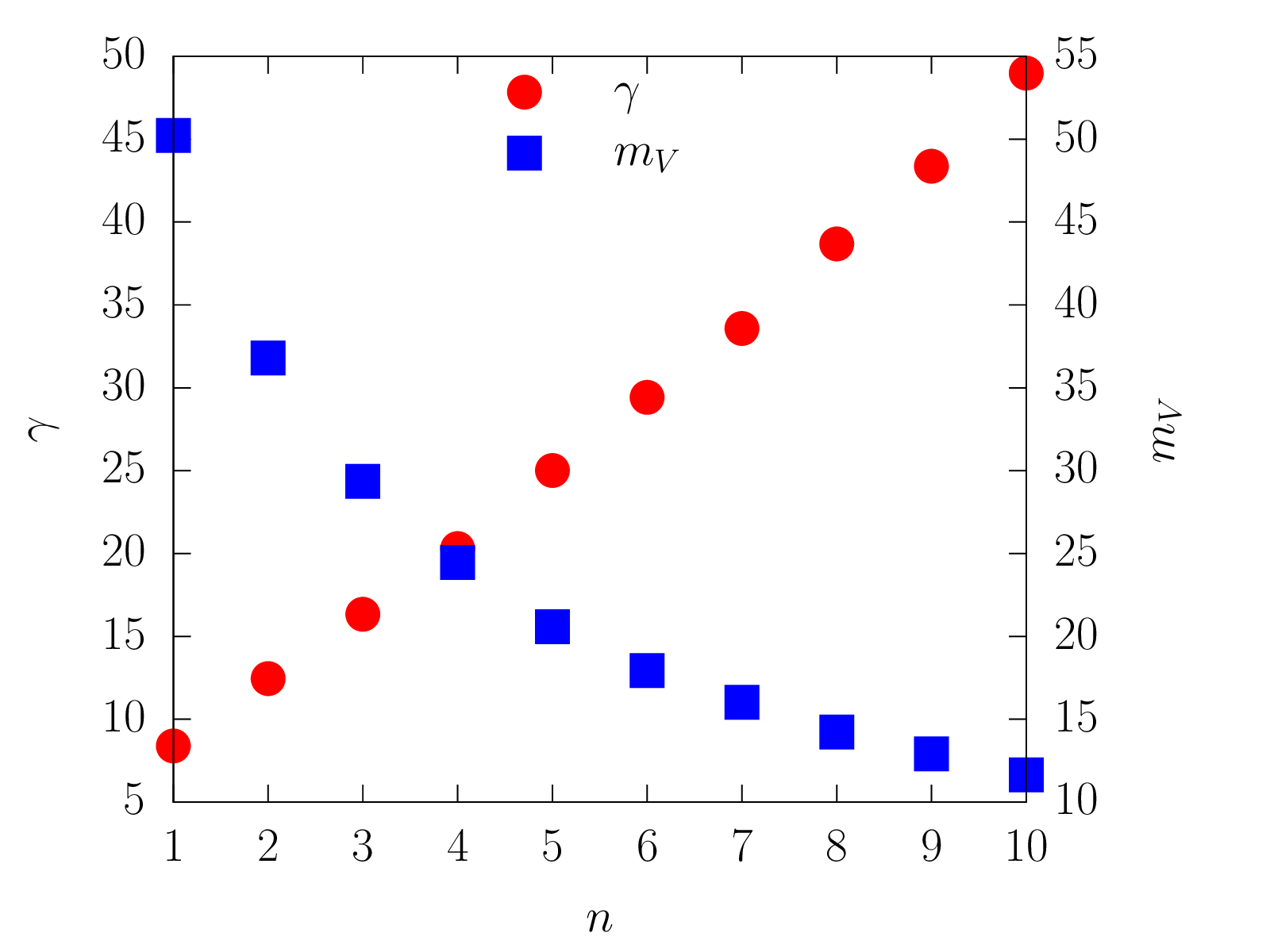}
 \caption{Scaling exponent $\gamma$ and isochoric fragility parameter $m_V$ as a function of side group length $n$. The computations are performed for the poly($\alpha$-olefin) structure with $z=6$,  $a_{cell}=2.9$ \AA{}, $\epsilon/k_B=200$ K, $E_b^b/k_B=560$ K, $E_b^s/k_B=100$ K and $N=8000$.}
\end{figure}

The side group length $n$ strongly affects the properties of polymers. The dependence of the glass transition temperature and the fragility on the side chain length has been extensively examined in previous works.~\cite{JacekACP2008, JacekJPCB2005a, JacekJPCB2005b, JacekJCP2005, JacekJCP2006} In particular, qualitatively different behaviors for the variation of the glass transition temperature with length of side chain has been revealed for different classes of polymers that are characterized by the ratio of side group to backbone bending energies. Thus, the generalized entropy theory establishes the side group length as a  powerful controllable tool for regulating the glass transition temperature and the fragility of glass-forming polymers. In a previous application of this theory, Stukalin \textit{et al.}~\cite{Evgeny} propose a set of molecular parameters to model glass formation of poly($\alpha$-olefins). We thus choose this model of poly($\alpha$-olefins) to illustrate the influence of side group length on the thermodynamic scaling of polymer melts. Figure 10 displays the computed variation of the scaling exponent and the isochoric fragility parameter with the side group length.  $\gamma$ increases with $n$, while $m_V$ again varies in the opposite direction, implying that a longer side chain leads to the enhancement of the density effect on the dynamics of poly($\alpha$-olefins). This is understandable since the side chains ($E_b^s/k_B=100$ K) are more flexible than the backbone ($E_b^b/k_B=560$ K) in this model, and thus a longer side chain improves the packing efficiency. This, in turn, exerts impacts on the scaling exponent and the fragility.

\subsection{Correlations between the scaling exponent and measures of fragility}

\begin{figure}[tb]
 \centering
 \includegraphics[angle=0,width=0.48\textwidth]{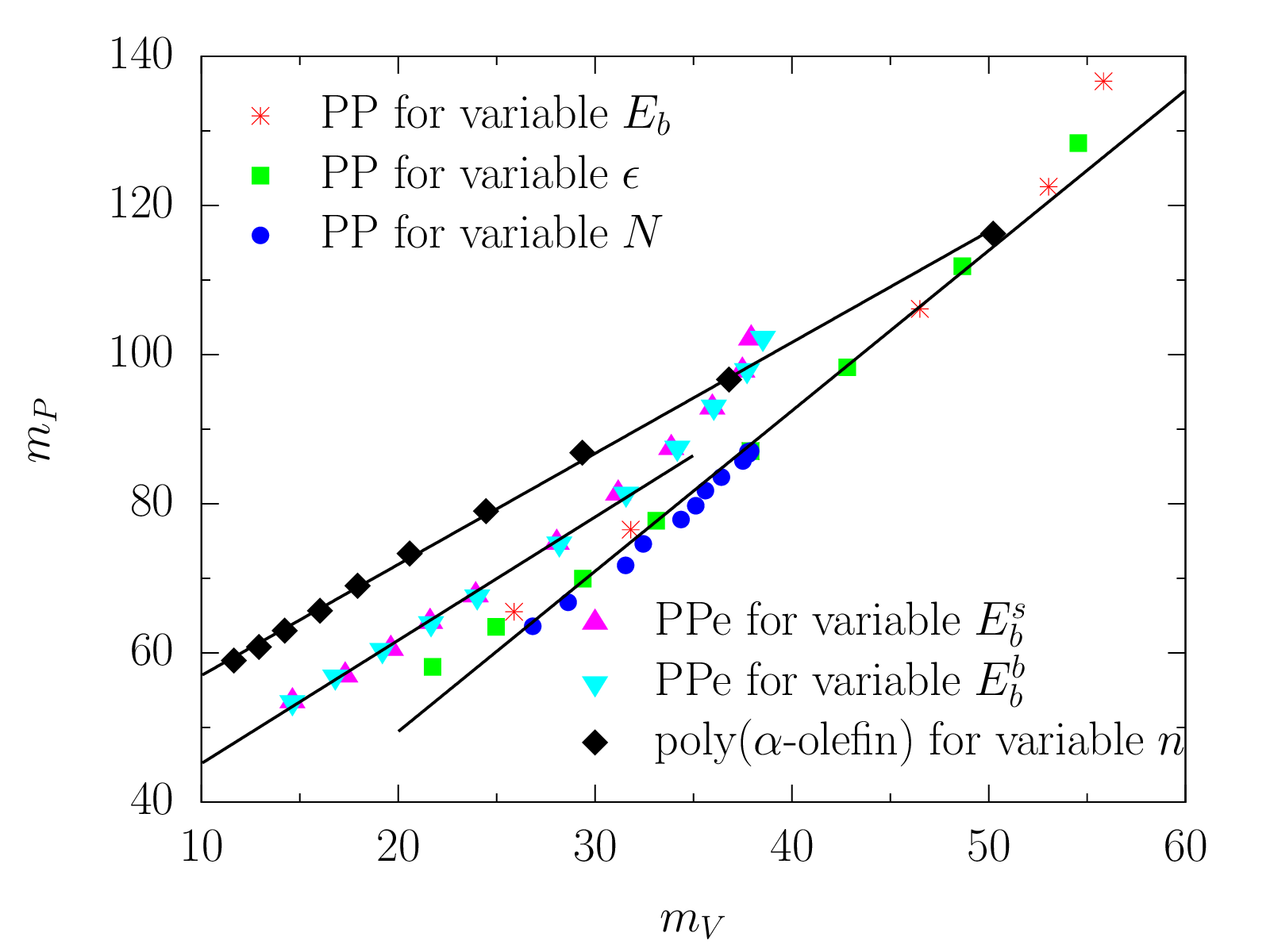}
 \caption{Correlations between isobaric fragility parameter $m_P$ at $P=0.1$ MPa and isochoric fragility parameter $m_V$ when varying individual molecular parameters. The lines are a guide to the eye.}
\end{figure}

\begin{figure}[b]
 \centering
 \includegraphics[angle=0,width=0.48\textwidth]{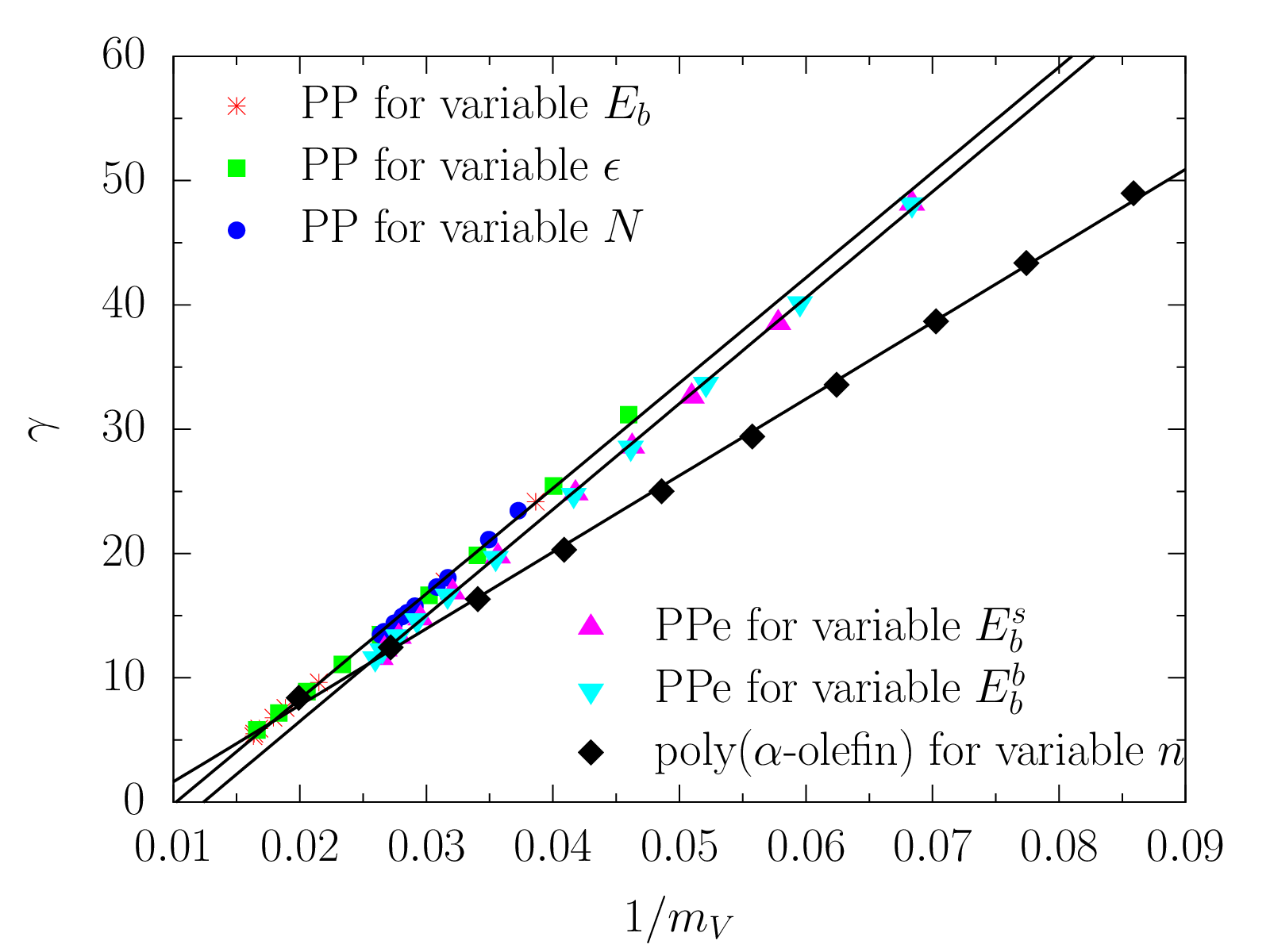}
 \caption{Correlations between scaling exponent $\gamma$ and isochoric fragility parameter $m_V$ when varying individual molecular parameters. The lines are linear fits according to Eq. (6) with the fitting parameters, $\gamma_0=-8.65$ and $b=847$ for chains with the poly(propylene) structure, $\gamma_0=-10.54$ and $b=852$ for poly(1-pentene), $\gamma_0=-4.52$ and $b=616$ for poly($\alpha$-olefin).}
\end{figure}

All of our computations indicate a negative correlation between the scaling exponent and the fragility, regardless of which molecular parameter is varied. Such a result is intriguing since it enables establishing correlations between the scaling exponent and the fragility and, hence, a universal physical mechanism for the fragility and the density scaling for the dynamics of polymer melts. Actually, the existence of correlations between the scaling exponent and the fragility has  previously been realized empirically.~\cite{RolandPRE2004, RolandPRE2005, CasaliniJNCS2006, GrzybowskiPRE2006} In particular, two empirical correlations have been suggested by Casalini and Roland~\cite{RolandPRE2005},
\begin{equation}
m_P=m_0+am_V,
\end{equation}
\begin{equation}
\gamma=\gamma_0+b/m_V,
\end{equation}
where $m_0$, $a$, $\gamma_0$ and $b$ are adjustable parameters. Since $m_P$ is dependent on pressure and $m_V$ is not, the coefficients in Eq. (5) vary with pressure, as noted previously.~\cite{RolandPRE2005, GrzybowskiPRE2006} Casalini and Roland~\cite{RolandPRE2005} indicate that these two correlations work well for most glass formers.

While these correlations are very appealing and have important consequences, no prior theory provides an explanation for the correlations. Our study provides the opportunity for testing the correlations, and, perhaps more importantly, for investigating how $m_P$, $m_V$ and $\gamma$ mutually correlate when a single molecular parameter is varied. Figures 11 and 12 display the $m_P-m_V$ and $\gamma-m_V^{-1}$ correlations, respectively, for varying molecular parameters. Remarkably, all of the data never fall onto a single master curve, but they can be roughly categorized into three groups. Choosing the PP structure for illustration, $m_P$ grows linearly with $m_V$ only in a restricted regime of lower values, while the linear relation between $\gamma$ and $m_V^{-1}$ appears perfect over the whole regime. Varying the bending energy, cohesive energy and chain length produces a single universal linear curve. Similar results are found for the structure of PPe. Finally, a linear relation applies for both $m_P-m_V$ and $\gamma-m_V^{-1}$ correlations when only the side group length is altered. Although a single universal relation cannot cover all cases, our computations clearly indicate the presence of correlations between the scaling exponent and fragility. Hence, our study points towards a universal understanding of the fragility and the density scaling of dynamics in polymers.

An important contribution of the generalized entropy theory is the identification of the packing efficiency as the main source of variation in the fragility of glass-forming polymers.~\cite{FreedACR2011, JacekACP2008} The packing efficiency in turn depends on chain rigidity, cohesive energy and molecular details, such as the backbone and side group lengths. Strong support for this simple picture for the origin of variations in fragility is provided by recent experiments~\cite{SokolovMac2008} and simulations.~\cite{PabloPRL2006, SokolovPCCP2013} These findings yield the immediate recognition that the effect of density on the dynamics of polymers, i.e., the magnitude of the scaling exponent, is also determined by the packing efficiency because of the strong correlations between the fragility and the scaling exponent. Hence, the influence of the bending energy, cohesive energy, backbone and side group lengths on the thermodynamic scaling can be rationalized in terms of the efficiency of packing of polymers in the melt.

\section{Summary}

We have employed the generalized entropy theory to investigate the influence of different molecular factors on an intriguing property, namely, the thermodynamic scaling of glass-forming polymers. Thermodynamic scaling emerges from the theory for relatively low pressures in this molecular theory of polymer glass formation. In particular, we systematically examine how the scaling exponent varies when changing a single molecular parameter, including the chain rigidity, the cohesive energy, the chain length and the side group length. All of these molecular factors have been shown to strongly affect the scaling exponent and hence the density effect on the dynamics of polymers. We explore correlations between the scaling exponent and the fragility. Instead of falling onto a single master curve, these correlations appears to depend on the molecular structure. Based on these correlations, we propose that the packing efficiency determines the magnitude of the scaling exponent and hence the relative importance of density for the dynamics of polymers, just as the packing efficiency constitutes a general determinant of fragility.

\begin{acknowledgments}
This work is supported by  the U.S. Department of Energy, Office of Basic Energy Sciences, Division of Materials Sciences and Engineering under Award DE-SC0008631. We are grateful to Jacek Dudowicz and Jack F. Douglas for several useful discussions.
\end{acknowledgments}



\begin{thebibliography}{0}
\bibitem{footnote1}The real-unit specific volume $v$ is calculated by $(1/\phi)(a_{cell}^3rN_{Av}/M_{mol})$, where $r$ denotes the number of lattice sites occupied by a single monomer, $M_{mol}$ is its molar mass and $N_{Av}$ is Avogadro's number. For the PP structure, $r=3$ and $M_{mol}=42$ g/mol.
\end{thebibliography}

\end{document}